# *A study in* blood at pinch

Shantimoy Kar, Aritra Kar, Kaustav Chaudhury, Tapas Kumar Maiti, Suman Chakraborty

The complex fluidic nature of blood – though necessary to serve different physiological purposes – gives rise to daunting challenges in developing unified conceptual paradigm describing the underlying physics of blood at pinch, which may otherwise be essential for understanding various bio-technological processes demanding precise and efficient handling of blood samples. Intuitively, a blood-drop may be formed simply by dripping. However, the pinch-off dynamics leading to blood-drop-breakup is elusively more complex than what may be portrayed by any unique model depicting the underlying morpho-dynamics, as our study reveals. With blood samples, here we observe two distinctive modes of the breakup process. One mode corresponds to incessant collapsing of a liquid-neck, while in other mode formation and thinning of an extended long thread leads to the breakup and drop formation. We further show that these modes are respectively described by power law and exponential law based universal scaling dynamics, depicting the temporal evolution of the neck thickness of the blood filament. Our results are likely to bear far-reaching consequences in diverse applications, ranging from forensic sciences to droplet based microfluidic technology.

**Corresponding author:** Suman Chakraborty (suman@mech.iitkgp.ernet.in)

**Affiliations**
SK: Advanced Technology Development Centre, Indian Institute of Technology Kharagpur, Kharagpur-721302, India
AK, KC, SC: Department of Mechanical Engineering, Indian Institute of Technology Kharagpur, Kharagpur-721302, India
TKM: Department of Biotechnology, Indian Institute of Technology Kharagpur, Kharagpur-721302, India

**Contributions**
SC conceived and supervised the work. SK oversaw the experiments conducted. AK and SK performed the experiments. KC performed the theoretical analysis. SK and KC processed the results. KC, SK, SC wrote the manuscript. TKM and SC edited the paper.

# Introduction

A drop of blood is possibly one of the most valuable entities in medical treatment, diagnosis, and in pathogenesis [1]. It's worth is overwhelmingly appreciated in forensic investigations [2], even by *Sherlock Holmes* in his feat during *A Study in Scarlet* (by *Sir Arthur Conan Doyle*). Thanks to wide gamut of research endeavours, tremendous technological advancements in the above mentioned practices have taken place over the years [3]. Despite elusively varying diversities, the physics governing many of these applications is intrinsically guided by a common underlying fact that as soon as the external stretching forces on a blood thread exceed the interfacial tension, the thread tends to elongate, thereby forming a liquid neck. Under external forcing, the neck may further elongate into an almost cylindrical bridge. The bridge may subsequently become narrow enough, and break as a consequence of instability mechanisms that may be effectively dampened by the blood viscosity. The pinch off dynamics of blood, thus, is expected to be modulated by its overwhelmingly complicated rheology. Although rigorous strives have been undertaken in conceptualizing the pinch-off dynamics of fluids with various levels of fluidic complexities [4–8], consensus on blood pinch-off dynamics is rather scarce. Here, we present the flow physics of blood pinch-off, through simple experiments of falling or dripping pendent drop from a tube. We bring in perspectives from the dynamics of complex-fluids towards rationalizing the observations, in terms of corroborating scaling laws depicting the temporal evolution of the neck-width of a breaking blood filament. These results can be of particular significance in a plethora of applications ranging from forensic sciences to droplet based microfluidics.

From the viewpoint of practical interests, developing technologies for blood sample analysis has been the key source of motivation for many previous research endeavours. Despite the level of sophistication achieved so far in this regard, the usual practices are such that once a blood sample is used, the same cannot be used further. In this respect, scrutiny of the pinch-off modalities could be a potential strategy for analysis, where same blood sample can be used repeatedly in a simple dripping experiment. This is of immense significance in forensic investigations, where, the available blood sample at the site of crime is limited [2]. From a fundamental viewpoint as well, the scientific challenges in understanding the blood pinch-off modalities cannot be overlooked, as attributable to the overall manifestation of the intricate dynamics of the included red blood cells (RBCs), white blood corpuscles (WBCs), fibrinogens and other proteins that are suspended in the plasma matrix [9].

In general, a liquid-droplet can be formed from a continuous stream by various physical mechanisms. On the local length-scale at the sites of droplet emancipation, the formation and thinning of a liquid neck is a characteristic signature of the instigation of pinch-off process observed in all types of fluids [5,7,10,11]. From quantitative perspective, evolution of the neck diameter $(d_{neck})$ in the time remaining to pinch-off $(\tau)$ is the key focus in the observation of any pinch-off process. Fluidic complexities enter into the picture in deciding the relationship between $d_{neck}$ and $\tau$. Orientation, clustering and deformation of the suspended matters in the blood sample, as well as a combined viscous and elastic nature, impart several fluidic complexities towards analyzing blood at pinch. This may be further complicated by the age and health of the concerned subject [9]. As a consequence, any unified relationship between $d_{neck}$ and $\tau$ is not possible to be foreseen at the very outset. This necessitates a comprehensive investigation of the physical features occurring over the neck region of a breaking blood droplet, with an interplay of the complex rheological features governing the response to the incipient stresses.

**Results and discussion**

We conduct experiments on falling pendent blood drop from a tube. Essentially, blood is flown through a cylindrical tube open to atmosphere. Subsequently, the images of the free-falling blood drop are captured and analysed (cf. section Methods for details). We explore pinch-off modalities for different blood samples with varying ranges of haematocrit concentration, HCT (defined as the total volume of the red blood cells or packed cell volume with respect to total blood volume). The resulting phenomenological features are shown in Fig. 1. Specifically, we observe two possible modes of pinch-off, as shown in Fig. 1: (i) Incessant neck collapsing mode, and (ii) extended thread breakup mode. Please refer the supplementary movie [12] for comprehensive visualization of the pinch-off process.

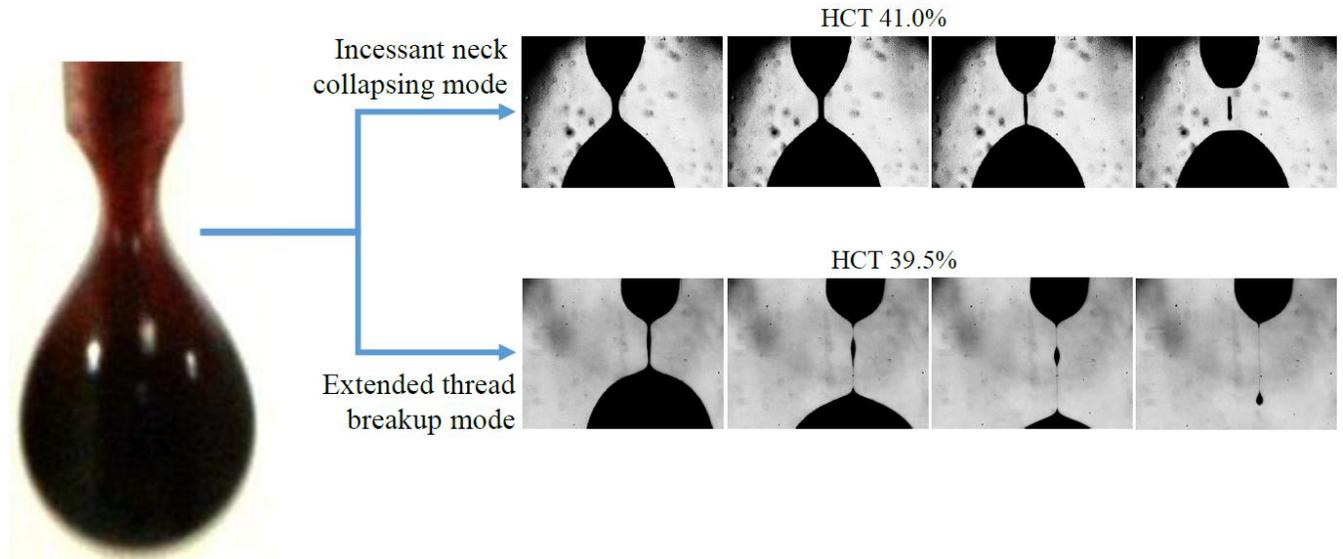

Fig. 1. Pinch-off of falling pendent drop of blood from tube. Two possible modes of pinch-off, as observed in the present experiments, are shown: (i) Incessant neck collapsing mode (shown for HCT 41.0%), and (ii) extended thread breakup mode (shown for HCT 39.5%). Dynamic visualization of the pinch-off processes are provided in the supplementary movie [12].

It appears tempting to explore the patterns of $d_{neck}$ evolution in $\tau$. However, prior to that, certain subtleties of the underlying physics at the site of pinch-off needs to be considered [10,11]. Fig. 2 highlights the zone of interest around the neck area. Here, the two-fluid interface assumes a curvature $\kappa \sim d_{neck}^{-1}$. Subsequently, a capillary pressure $p \sim \sigma/d_{neck}$ is developed within the liquid region, where $\sigma$ is the coefficient of surface tension at the two-fluid interface. Evidently, $p$ is higher at the narrowest $d_{neck}$ which is earmarked as the origin of the $r-z$ frame. As $d_{neck}$ reduces with $z$, a gradient $\nabla p \sim \sigma/(d_{neck} z)$ is developed along $z$, forcing the liquid within the neck to diverge away from the narrow zone, as shown in Fig. 2. This leads to subsequent neck thinning.

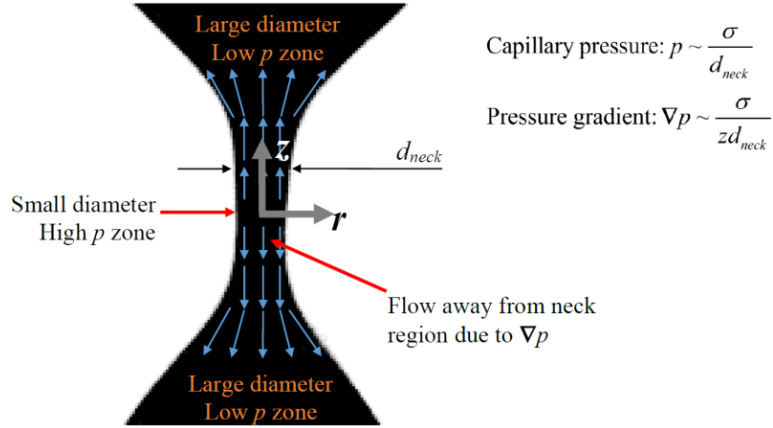

Fig. 2. Highlighted neck region with diameter $d_{neck}$ (from experimental observation), the zone of interest. Gradient of the capillary pressure drives the flow away from the narrowest part of the neck region (shown by the arrows), leading to emptying of liquid from the neck until breakup. The cylindrical coordinate system, as shown by $r-z$ coordinates, serves as the reference frame for theoretical descriptions. The azimuthal coordinate is on the plane with $z$ as the outward normal according to right handed system.

Detailed analysis over the neck region establishes that irrespective of the fluidic complexities, the predominant component of the velocity vector $\mathbf{u}$ is $u_z$ along $z$ direction [10,11,13]. Due to the incipient capillary forcing, the rate of change of momentum (per unit volume) of the liquid within the neck region $\rho \partial \mathbf{u}/\partial t + \rho \mathbf{u} \cdot \nabla \mathbf{u}$ scales as $\rho \partial \mathbf{u}/\partial t \sim \rho u_z / \tau$ (temporal inertia) and $\rho \mathbf{u} \cdot \nabla \mathbf{u} \sim \rho u_z^2 / z$ (convective inertia) over the characteristic length and time scales $z$ and $\tau$ respectively. Equivalence of these two contributors to inertia results in $u_z \sim z/\tau$, a scaling estimate for the spatio-temporal evolution of the velocity field. The fluidic complexities are reflected through the viscous contribution. We must note that blood viscosity may evolve nonlinearly with the strain rate, with asymptotic attainment of constant viscosity plateaus at both low and high strain rates [9].

One simple mathematical depiction of the constitutive behaviour of blood is the empirical power law formulation that generalizes Newton's law of viscosity to the form [14]: $\mathbf{T} = m\mathbf{S}^n + \mathbf{T}^*$, where $\mathbf{T}$ is a generalized stress dyadic, $\mathbf{S}$ is a generalized rate of strain dyadic, $m$ and $n$ are the flow consistency and the flow behaviour indices respectively, [15]. Here $\mathbf{T}^*$ is a generalized yield-stress dyadic, which defines the so-called Bingham plastic fluids, for which a minimum amount of force is necessary to initiate the flow. It has been established that human blood shows only very slight Bingham plastic property, as attributable to the presence of the protein fibrinogen. Reported results [14] indicate that the yield stress of blood is quite small (<<0.025 Pa), and therefore may be neglected under dynamical conditions. The indices $n$ and $m$ have a primary functional dependence on HCT, and certain plasma proteins – most notably, the globulins. In the non-linearly viscous regime, one may write: $\mathbf{T} = m\left[\nabla\mathbf{u}+(\nabla\mathbf{u})^T\right]^n$. Subsequently, the resulting viscous contribution scales as $\nabla\cdot\mathbf{T} \sim mu_z^n/z^{n+1}$. This type of nonlinear approximation for blood constitutive behaviour is a common consideration in the literature, where typically $n = 0.70 - 0.78$ for healthy human individuals [9,16].

Equating the contributions of inertia $(\rho\partial\mathbf{u}/\partial t)$, capillarity $(\nabla p)$ and viscous forcing $(\nabla\cdot\mathbf{T})$, along with the consideration $u_z \sim z/\tau$, scaling relationship between $d_{neck}$ and $\tau$ can be obtained. For the nonlinear viscosity regime, the scaling relation reads as $d_{neck} \sim \tau^n$. Evidently, the scaling relation over the Newtonian asymptotic limits can be obtained as $d_{neck} \sim \tau$, either by setting $n \to 1$ or using the equivalence of the above mentioned forcing factors with Newtonian viscous contribution. In general, the relation can be recast as $d_{neck} \sim \tau^\alpha$ where the power law exponent $\alpha$ bears the signature of the viscous contribution [4,7,10,15].

The evolution of neck diameter $d_{neck}$ in the time remaining to pinch-off $\tau$, for the blood samples employed in our study, are shown in Fig. 3. Existence of a similar feature near the pinch-off $(\tau \to 0)$ is noteworthy. The similarity is characterized by scaling relation of the form $d_{neck} \sim \tau^\alpha$ near pinch-off. Values of $\alpha$ corresponding to given HCTs are also indicated in the figure. Interestingly, the ranges of $\alpha$ presented in Fig. 3 are well within the window of $n$ (typically 0.70–0.78 [9,16]) for healthy individuals. It is worth mentioning that close convergence of the experimental data with the theoretical paradigm is noteworthy from the figure and thus validating the scaling arguments.

It is important to explore genesis of such HCT-dependent peculiar pinch-off characteristics. In effect, red blood cells (RBCs) form clusters at low shear rates, known as rouleaux [17]. In the absence of plasma proteins, RBCs would not show any kind of aggregation, as a consequence of repulsive electrostatic interactions between the cells due to the negative charges in the cell membranes [18]. However, in the presence of plasma proteins like fibrinogen and globulin in particular, RBCs tend to aggregate. It may be possible [19] that the tails of high molecular weight molecules tend to adhere to the adjacent cells, in the process linking them together over distances that ensure the adhesive forces to be dominating over the repulsive ones. This, in turn, increases the effective volume of RBCs, so as to enhance the effective viscosity of blood. With increase in HCT, there is an effective increment in the RBC volume, bringing in higher level of energy dissipation in the flow. However, upon increase in the shear rate at the necking region, the aggregates tend to break up, resulting in a dramatic decrement in the resistive shear, with a delicate interplay between the conflicting effects of the increase of HCT that attempts to enhance the blood viscosity on one hand, and a shear-induced reduction in the cell size that attempts to decrease the same on the other. In addition, at high shear rates, the RBCs acquire a unique capability to adapt to a new shape in response to the deforming forces, by promoting their orientations to the flow streamlines, thereby reducing viscous drag.

We must note that the above mentioned scaling argument $d_{neck} \sim \tau^n$ is for pinch-off by incessant neck collapsing. The other mode – i.e. pinch-off by extended thread breakup – does not conform to such power law based scaling argument. Thus, in Fig. 3, we presented the results only for pinch-off by incessant neck collapsing. The physics of the other mode of blood pinch-off may be appreciated by first noting that the capillary pressure driven flow causes divergence of the liquid from the narrow neck region. Subsequently, the neck collapses due to emptying of liquid from that region. However, in due course, the entities suspended in blood experience stretching due to the diverging flow within the neck. Owing to the elastic nature of suspended entities, the neck region behaves like an elastic rod under extension leading to the formation of an elongated slender liquid column, prior to breakup. This type of pinch-off by extended thread breakup is observed for polymer solutions (having elastic particles suspended within a liquid medium) [8,13,20].

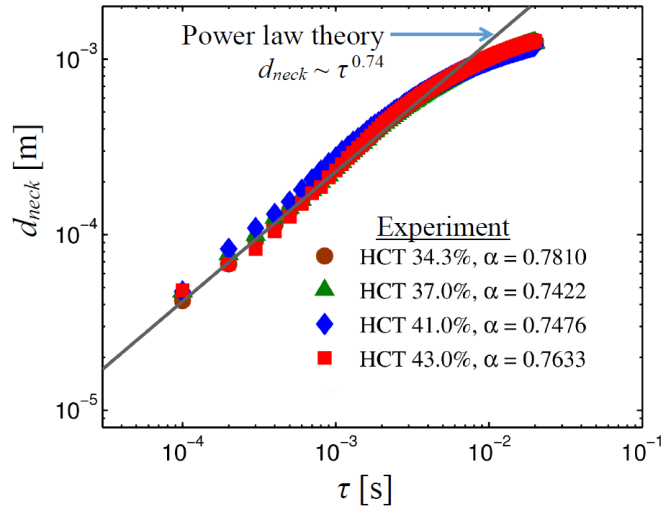

Fig. 3. Evolution of neck diameter $d_{neck}$ with the time remaining to pinch-off $\tau$, for blood samples undergoing pinch-off at different HCT. Maximum standard deviation for the experimental results is $1.1346 \times 10^{-4}$ m. At each HCT, the experimental data are fit with the scaling $d_{neck} \sim \tau^\alpha$ near pinch-off. The corresponding best fit values of $\alpha$ are indicated against each HCT value.

The stretching or extension can be characterized by the relaxation time $T_R$ of the suspended elastic macromolecules [20], related to the extensional strain rate as $1/3T_R = -d_{neck}^{-1} \partial d_{neck}/\partial t$. The negative sign is due to decrease in $d_{neck}$ with $t$. In terms of the time remaining to pinchoff $\tau = t_p - t$ (with $t_p$ being the time of pinch-off), the relation reads as $1/3T_R = d_{neck}^{-1} \partial d_{neck}/\partial \tau$. This relation describes $d_{neck} \sim e^{\tau/3T_R}$, an exponential evolution of $d_{neck}$ in $\tau$, under extensional stretching.

This is in contrast to the power law based evolution $d_{neck} \sim \tau^n$, during incessant collapsing. Thus, pinch-off by 'incessant neck collapsing' and 'extended thread breakup' are distinguished, respectively, by the 'power law' and 'exponential law' of the neck-diameter evolution. In Fig. 4 we show the evolution of $d_{neck}$ in $\tau$ at various HCT levels for both the modes of breakup. In the figure we also present the exponential $d_{neck} \sim e^{\tau/3T_R}$ and power law $d_{neck} \sim \tau^n$ (with $n = 0.74$) based theoretical paradigm.

The HCT 41% data presented in Fig. 4 corresponds to pinch-off by incessant neck collapsing. Thus, it agrees closely to $d_{neck} \sim \tau^{0.74}$ than $d_{neck} \sim e^{\tau/3T_R}$. The data for HCT 39.5%, on the other hand, is reminiscent of pinch-off by extended thread breakup. In this case formation of prominent long thread is observed, as highlighted in Fig. 4. Accordingly, this situation corresponds to exponential evolution $d_{neck} \sim e^{\tau/3T_R}$. The data for HCT 32.3%, 44.8% and 50.9% at first sight appears to follow pinch-off by incessant neck collapsing, as highlighted in Fig. 4. However, those data agree well with the exponential law, as evident from Fig. 4. Thus, it appears that extensional mode of breakup is dominant here rather than incessant neck collapsing. Further elucidation can be obtained from the predicted values of $T_R$, as indicated in Fig. 4 against the exponential law based conceptualization of fitting the experimental data. From the figure, it is evident that increase in the relaxation time $T_R$ instigates extensional mode of breakup.

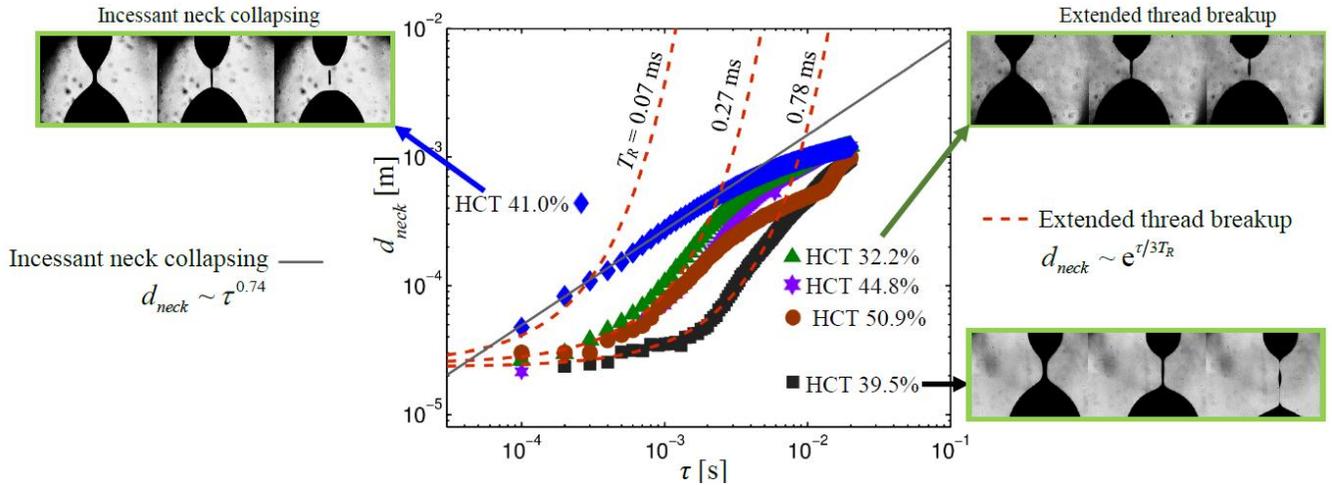

Fig. 4. Evolution of neck diameter $d_{neck}$ in time remaining to pinchoff $\tau$ at various HCT levels: Involvement of extensional mode during breakup. Standard deviations for each data set are as follows: HCT 39.5% = 3.68×10⁻⁴ m, HCT 44.8% = 3.18×10⁻⁴ m, HCT 32.2% = 3×10⁻⁴ m, HCT 50.9% = 6.98×10⁻⁴ m, HCT 41.0% = 3.04×10⁻⁵ m. A few snapshots around the neck region near pinch-off are also presented, for comprehensiveness. The extensional mode is described by the exponential evolution of neck diameter $d_{neck} \sim e^{\tau/3T_R}$ (dashed lines) where $T_R$ is the relaxation time for extension of the suspended elastic entities in blood. Estimated $T_R$ values, based on best fit of the experimental data, are indicated against each exponential observation. Data for pinch-off by neck collapsing (for HCT 41.0%) are also presented for the sake of comparison along with its power law based paradigm of neck collapsing $d_{neck} \sim \tau^{0.74}$ (solid line).

To highlight on the aspect of relaxation, we bring in perspective from 'bead-spring-dumbbell' model for rationalizing constitutive behaviour of a complex-liquid from microscopic point of view [21,22]. The model considers a dumbbell like situation where two spheres of radius $a_p$ are separated by a distance $\mathbf{L}$ (vector form). The spheres are acted upon by drag force due to fluid flow having effective viscosity $\mu_{eff}$ which tries to extend the length. However, a spring-like recoiling force with stiffness $k_s$ tries to resists such elongation. Force balance analysis for the dumbbell results in the relation [21,22] $3T_R \dot{\mathbf{L}} = -\mathbf{L} + 3T_R \mathbf{L} \cdot \nabla \mathbf{u}$, where the relaxation time can be conceptualized as $T_R = 2\pi \mu_{eff} a_p / k_s$. Here, the over dot represents the derivative with respect to time. Evidently, for very small $T_R$, the separation length is almost zero, i.e. the spheres are attached together strongly. With progressive increase in $T_R$, the elongation starts to be prominent.

The mechanism depicted as above can be related to blood rheology by noting that the plasma proteins remain attached over the surfaces of the RBCs, acting as the binder to form RBC cluster. Thus, a 'bead-spring-dumbbell' like structures can be considered to exist in blood where the plasma proteins act as the spring. When the binding force is strong enough (or equivalently high $k_s$ scenario), $T_R$ is very small. Thus, the clusters would transport with the flow as discrete entities. In the presence of strong shear, the cluster formation may be impeded, resulting in the flow of separate RBCs. Both these situations pertain to viscous behaviour of the fluid over the neck region, reflected through the pinch-off by incessant neck collapsing mode. However, the elevation in $T_R$ due to smaller spring force (smaller $k_s$) results in stretched spring-like extension. Subsequently, pinch-off by extended thread breakup can be observed. It appears that it is essentially the connection between the RBCs which decides the extensional mode. In particular, the spring-like behaviour of the plasma proteins that binds the RBCs together is the key factor here.

## Conclusions

Summarily, pinch-off of blood can take place either by incessant neck collapsing or by extended thread breakup. The former mode is characterized by the emptying of liquid from the neck region, leading to a power law based collapsing of the neck in the time remaining to pinch-off. The later mode has it genesis in the extensional relaxation of the RBCs' connection mediated through plasma proteins. The extensional mode pertains to an exponential evolution of neck diameter in the time remaining to pinch-off. Depending on the arrangement of the suspended entities in blood, a blood sample can exhibit either of the mentioned modes of breakup. Despite elusively varying age and health related diversities, the macroscopic manifestation of the underlying phenomenon is essentially governed by an inhomogeneous distribution of corpuscles in the blood sample which itself changes with time. Following this conjecture, our generalized depiction of the morphodynamics of the blood samples at pinch, considering the two modes of break-up unveiled by our experiments, may bear immense consequences in diverse applications ranging from forensic sciences to droplet based bio-microfluidics.

## Methods

Experiments are performed as per the Institutional ethical guidelines of the authors (Approval No. IIT/SRIC/AR/2012),. Informed consents are taken while collecting the blood samples from the volunteers.

**Sample preparation**

The collected blood samples are preserved in anti-coagulant coated vials at around 20°C. The samples are directly used for experiments without any dilution or chemical treatment. The objective is to maintain the physiologically pertinent condition to the extent possible. Subsequently, experiments are executed within 24 hours after collection of the sample. We explore the experimentation for different blood samples for varying ranges of HCT.

**Experimental observation of blood pinch-off**

Fig. M1 schematically shows the present experimental consideration. The blood is flown through a syringe pump (HARVARD APARATUS PHD2000) through a tube having diameter of 2 mm. One end of the tube is inserted through the top surface of an enclosed glass tube. It is known that the physical subtleties over the characteristic length and time scales, around the neck area during pinch-off, are independent of the flow features at the far regions [10]. Thus, volume flow rate has no influence on the pinch-off dynamics. However, it is usual to employ a constant flow rate so as to ensure intermittent supply of liquid [5]. Here, we maintain a volume flow rate of $50\mu l/\min$, through the syringe pump.

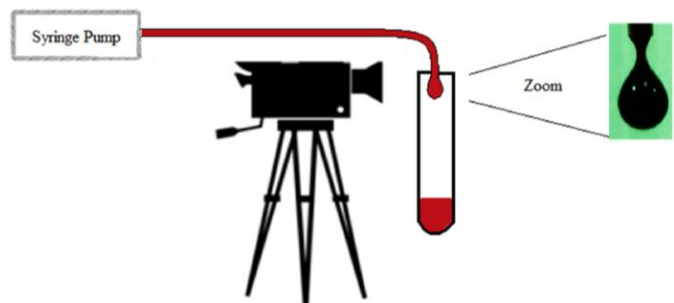

Fig. M1. Schematic illustration of the experimental setup

Blood pinches off at the end of the tube in a surrounding medium of air (it can be considered as free falling blood drop in an enclosed glass tube) (cf. magnified portion of Fig. M1). The images of the free-falling blood drop are captured through a high-speed camera (AMETEK V641) at a speed of 10000 fps with a resolution of (640 pixels X 480 pixels). All the captured images are then post-processed through an in-house MATLAB© code which takes care about the calculation of neck radius using edge detection method. Repeated experiments are performed, and the average behaviours along with the corresponding standard deviations are reported.

**Acknowledgement**

The authors gratefully acknowledge the financial support provided by the Indian Institute of Technology Kharagpur, India [Sanction Letter no.: IIT/SRIC/ATDC/CEM/2013-14/118, dated 19.12.2013]. SK acknowledges the fellowship from Council of Scientific and Industrial Research (CSIR) for carrying out the research work.